\documentstyle[epsfig,amsmath,12pt]{article}

\author{Petr Z\'avada\\
\\\
{\it Institute of Physics, Academy of Sciences of Czech Republic}\\
{\it Na Slovance 2, CZ-180 40 Prague 8}\\
E-mail: zavada@fzu.cz}
\title{Proton spin structure in the rest frame
}
\date{August 5, 1997
}
\begin{document}

\maketitle
\begin {center}
(To be published in Phys. Rev. D)
\end{center}
\begin{abstract}
It is shown that the quark-parton model in the standard infinite momentum
approach overestimates the proton spin structure function $g_1(x)$ in
comparison with the approach taking consistently into account the internal
motion of quarks described by a spherical phase space in the proton rest
frame. Particularly, it is shown the first moment of the spin structure
function in the latter approach, assuming only the valence quarks
contribution to the proton spin, does not contradict to the experimental
data.
\end{abstract}

\section{Introduction}

The proton spin problem attracting significant attention during last few
years was triggered by the surprising results \cite{emc} of the European
Muon Collaboration (EMC), which analyzed data on the polarized deep
inelastic scattering (DIS). Since that time the hundreds of papers have been
devoted to this topic, for the present status see e.g. \cite{spin}, \cite
{dis} and the comprehensive overview \cite{hai}.

The essence of the problem is the following. From the very natural
assumption, that proton spin is created by the composition of the spins of
three valence quarks being in $s$-state, one can estimate value of the first
moment $\Gamma _1^p$ of the spin structure function $g_1^p(x)$%
\begin{equation}
\label{i1}\Gamma _1^p=\int_0^1g_1^p(x)dx\simeq 0.17. 
\end{equation}
In fact, such value was well reproduced in the SLAC experiment preceding the
EMC. Nevertheless, the EMC covering also a lower $x$ region, has
convincingly shown, the first moment is considerably lower: $\Gamma
_1^p=0.126\pm 0.18.$ And the latter experiments \cite{smc}, \cite{e143} gave
values compatible with the EMC. Such values can hardly correspond with the
concept that proton spin is a simple sum of the valence spins. In fact, a
global fit \cite{qcd} to all available data evaluated at a common $Q^2$ in a
consistent treatment of higher-order pertubative QCD effects suggests for
the spin carried by the quarks the value less than one third the proton
spin. So there is a question, what is the proton spin made?

In this paper we discuss the spin structure functions in the approach \cite
{zav} based on the proton rest frame and make a comparison with the standard
approach based on the infinite momentum frame (IMF). We do not attempt to
account for all the details important for the complete description of the
polarized proton, as e.g. the constrains resulting from axial vector current
operators, but we rather try to isolate the net effect of the oversimplified
kinematics in the IMF picture. Since this paper should be read together with 
\cite{zav}, for convenience we refer to the equations and figures in the
previous paper simply with prefix 'P', e.g. see Eq. (P3.41).

\section{Spin structure functions}

In our paper \cite{zav} the master equation (P3.41) has been based on the
standard symmetric tensors (P3.33) and (P3.34) corresponding to the
unpolarized DIS. After introduction the spin terms into both the tensors
(see e.g. \cite{fey}, Eqs. (33.9), (33.10)) our spin equation reads%
$$
P_\alpha P_\beta \frac{W_2}{M^2}-g_{\alpha \beta }W_1+i\epsilon _{\alpha
\beta \lambda \sigma }q^\lambda \left[ s^\sigma MG_1+(Pqs^\sigma -sqP^\sigma
)\frac{G_2}M\right] +A(P_\alpha q_\beta +P_\beta q_\alpha )+Bq_\alpha
q_\beta 
$$
$$
=\frac{P_0}M\int \frac{G(p)}{p_0}(2p_\alpha p_\beta -g_{\alpha \beta
}\,pq)\delta ((p+q)^2-m^2)d^3p 
$$

\begin{equation}
\label{ss1}+\frac{P_0}M\int \frac{H(p)}{p_0}i\epsilon _{\alpha \beta \lambda
\sigma }q^\lambda mw^\sigma \delta ((p+q)^2-m^2)d^3p, 
\end{equation}
where $G$ and $H$ are related to the polarized quark distributions 
\begin{equation}
\label{ss2}G(p)=\sum_je_j^2(h_j^{\uparrow }(p)+h_j^{\downarrow }(p)), 
\end{equation}
\begin{equation}
\label{ss3}H(p)=\sum_je_j^2(h_j^{\uparrow }(p)-h_j^{\downarrow }(p)) 
\end{equation}
and the spin fourvectors fulfill 
\begin{equation}
\label{ss4}s_\mu s^\mu =w_\mu w^\mu =-1,\qquad s_\mu P^\mu =w_\mu p^\mu =0. 
\end{equation}
The Eq. (\ref{ss1}) requires for the spin terms 
\begin{equation}
\label{ss5}s^\sigma MG_1+(Pqs^\sigma -sqP^\sigma )\frac{G_2}M=\frac m{2M\nu
}\int \frac{H(p)}{p_0}w^\sigma \delta \left( \frac{pq}{M\nu }-x\right) d^3p, 
\end{equation}
where we use for the $\delta -$function the relation (P3.46).

Now, to be more definite, let us consider a simple scenario assuming the
following.

\noindent 1) To the function $H$ in Eq. (\ref{ss3}) only the valence quarks
contribute.

\noindent 2) In the proton rest frame the valence quarks are in the $s-$%
state and their momenta distributions have the same (spherically symmetric)
shape for $u$ and $d$ quarks 
\begin{equation}
\label{ss6}h_d(p)=\frac 12h_u(p)\equiv h(p). 
\end{equation}
3) Both the quarks have the same effective mass $m^2=p^2$ in the sense
suggested in \cite{zav}. In this way it is assumed the effective mass of the
valence quark is characterized by the one fixed value, on the end this point
will obtain more realistic form .

\noindent 4) All the three quarks contribute to the proton spin equally 
\begin{equation}
\label{ss7}h_d^{\uparrow }-h_d^{\downarrow }=\frac 12(h_u^{\uparrow
}-h_u^{\downarrow })\equiv \Delta h(p_0)=\frac 13h(p_0),\qquad p_0=\sqrt{%
m^2+p_1^2+p_2^2+p_3^2}. 
\end{equation}
Since the proton and each of the three quarks have spin one half, spin of
two quarks must cancel and spin of the third, remaining, gives the proton
spin, so the last equation implies 
\begin{equation}
\label{ss8}3\int \Delta h(p_0)d^3p=1. 
\end{equation}
The combination with (\ref{ss3}) gives 
\begin{equation}
\label{ss9}H(p_0)=2\frac 49\Delta h(p_0)+\frac 19\Delta h(p_0)=\Delta h(p_0) 
\end{equation}
and 
\begin{equation}
\label{ss10}\int H(p_0)d^3p=\frac 13. 
\end{equation}

Now, let us assume the proton is polarized in the direction of the collision
axis (coordinate one), then Eq. (\ref{ss4}) requires for the proton at rest 
\begin{equation}
\label{ss11}s=(0,1,0,0) 
\end{equation}
and for the quark with fourmomentum $p$%
\begin{equation}
\label{ss12}w=\left( \frac{p_1}{\sqrt{p_0^2-p_1^2}},\frac{p_0}{\sqrt{%
p_0^2-p_1^2}},0,0\right) . 
\end{equation}
The contracting of Eq. (\ref{ss5}) with $P_\sigma $ and $s_\sigma $ (or
equivalently, simply taking $\sigma =0,1)$ gives the equations 
\begin{equation}
\label{ss13}q_1G_2=\frac m{2M\nu }\int \frac{H(p_0)}{p_0}\frac{p_1}{\sqrt{%
p_0^2-p_1^2}}\delta \left( \frac{pq}{M\nu }-x\right) d^3p, 
\end{equation}
\begin{equation}
\label{ss14}MG_1+\nu G_2=\frac m{2M\nu }\int \frac{H(p_0)}{p_0}\frac{p_0}{
\sqrt{p_0^2-p_1^2}}\delta \left( \frac{pq}{M\nu }-x\right) d^3p. 
\end{equation}
In the next step we apply the approximations from the Eqs. (P2.9) and
(P2.14) 
\begin{equation}
\label{ss15}q_1\simeq -\nu ,\qquad \frac{pq}{M\nu }\simeq \frac{p_0+p_1}M. 
\end{equation}
Let us note, the negative sign in the first relation is connected with the
choice of the lepton beam direction giving the Eq. (P2.14). The opposite
choice should give 
\begin{equation}
\label{ss16}q_1\simeq +\nu ,\qquad \frac{pq}{M\nu }\simeq \frac{p_0-p_1}M 
\end{equation}
and one can check the both alternatives result in the equal pairs $G_1,G_2$,
which reads 
\begin{equation}
\label{ss17}2g_1(x)\equiv 2M^2\nu G_1=m\int \frac{H(p_0)}{p_0}\frac{p_0+p_1}{
\sqrt{p_0^2-p_1^2}}\delta \left( \frac{p_0+p_1}M-x\right) d^3p, 
\end{equation}
\begin{equation}
\label{ss18}2g_2(x)\equiv 2M\nu ^2G_2=-m\int \frac{H(p_0)}{p_0}\frac{p_1}{
\sqrt{p_0^2-p_1^2}}\delta \left( \frac{p_0+p_1}M-x\right) d^3p. 
\end{equation}
Let us remark the integration of Eqs. (\ref{ss13}) and (\ref{ss18}) over $x$
gives on r.h.s. the integral 
\begin{equation}
\label{ssa18}\int \frac{H(p_0)}{p_0}\frac{p_1}{\sqrt{p_0^2-p_1^2}}d^3p=0, 
\end{equation}
which is zero due to spherical symmetry. Therefore in this approach the
first moment of $g_2(x)$ is zero as well. In the next we shall pay attention
particularly to the function $g_1$, which can be rewritten 
\begin{equation}
\label{ss19}2g_1(x)=\frac{x_0}3\int h(p_0)\frac M{p_0}\sqrt{\frac{p_0+p_1}{%
p_0-p_1}}\delta \left( \frac{p_0+p_1}M-x\right) d^3p,\qquad x_0=\frac mM. 
\end{equation}
What our assumptions 1)-4) do mean in the language of the standard IMF
approach? In \cite{zav} (end of section III.B) we have shown our approach is
equivalent to the standard one for the static quarks described by the
distribution function $h(p_0)$ sharply peaked around $m$. In such a case the
last equation for $p_0\approx m,$ $p_1\approx 0$ after combining with (\ref
{ss3}) and (P3.1) gives 
\begin{equation}
\label{ss20}2g_1(x)=\int \sum_je_j^2(h_j^{\uparrow }(p_0)-h_j^{\downarrow
}(p_0))\delta \left( \frac{p_0+p_1}M-x\right) d^3p=\sum_je_j^2(f_j^{\uparrow
}(x)-f_j^{\downarrow }(x)), 
\end{equation}
where $f_j(x)$ are corresponding distribution functions in the IMF, so in
this limiting case our spin equation (\ref{ss19}) is also identical with the
standard one, see Eq. (33.14) in \cite{fey}. On the other hand the last
equation can be in our simplified scenario rewritten 
\begin{equation}
\label{ss21}2g_1(x)=\frac 13\int h(p_0)\delta \left( \frac{p_0+p_1}%
M-x\right) d^3p=\frac 13f(x)=\frac{F_{2val}(x)}{3x}. 
\end{equation}
This relation could be roughly expected in the standard IMF approach and
correspondingly 
\begin{equation}
\label{ss22}\Gamma _{IMF}\equiv \int g_1(x)dx=\frac 16\int f(x)dx=\frac 16. 
\end{equation}
The Eqs. (\ref{ss19}) and (\ref{ss21}) are equivalent for the static quarks,
but how they differ for the non static ones? In accordance with (P3.54) let
us denote 
\begin{equation}
\label{ss23}V_j(x)\equiv \int h(p_0)\left( \frac{p_0}M\right) ^j\delta
\left( \frac{p_0+p_1}M-x\right) d^3p, 
\end{equation}
then (P3.52) and (\ref{ss21}) give 
\begin{equation}
\label{ss24}2g_1(x)=\frac{xV_{-1}(x)}3,\qquad \Gamma _{IMF}=\frac 16\int
xV_{-1}(x)dx. 
\end{equation}
Now, let us calculate the corresponding integral from our rest frame
equation (\ref{ss19}) 
\begin{equation}
\label{ss25}\Gamma _{lab}=\frac{x_0}6\int \int h(p_0)\frac M{p_0}\sqrt{\frac{%
p_0+p_1}{p_0-p_1}}\delta \left( \frac{p_0+p_1}M-x\right) d^3pdx. 
\end{equation}
Due to the $\delta -$ function, the square root term in the integral can be
rewritten 
\begin{equation}
\label{ss26}\sqrt{\frac{p_0+p_1}{p_0-p_1}}=\sqrt{\frac{Mx}{2p_0-Mx}}=\sqrt{
\frac{Mx}{2p_0}}\left( 1-\frac{Mx}{2p_0}\right) ^{-\frac 12}=\left( \frac{Mx
}{2p_0}\right) ^{\frac 12}\sum_{j=0}^\infty {{-\frac 12}\choose j}(-1)^j\left( 
\frac{Mx}{2p_0}\right) ^j
\end{equation}
and with the using of (\ref{ss23}) the integral correspondingly 
\begin{equation}
\label{ss27}\Gamma _{lab}=\frac{x_0}6\int \sum_{j=0}^\infty {{-\frac 12}%
\choose j}(-1)^jV_{-j-3/2}(x)\left( \frac x2\right) ^{j+1/2}dx.
\end{equation}
The integration by parts combined with the relations (P3.56) gives%
$$
\int V_{-j-3/2}(x)\left( \frac x2\right) ^{j+1/2}dx=\int V_{-j-3/2}^{\prime
}(x)\frac{2\left( x/2\right) ^{j+3/2}}{j+3/2}dx=\int V_0^{\prime }(x)\left(
\frac x2+\frac{x_0^2}{2x}\right) ^{-j-3/2}\frac{2\left( x/2\right) ^{j+3/2}}{%
j+3/2}dx 
$$
$$
=\int V_0^{\prime }(x)\frac 2{j+3/2}\left( \frac 1{1+x_0^2/x^2}\right)
^{j+3/2}dx=\int V_0(x)2\left( \frac 1{1+x_0^2/x^2}\right) ^{j+1/2}\frac{%
2x_0^2/x^3}{\left( 1+x_0^2/x^2\right) ^2}dx. 
$$
If we denote $t\equiv x_0^2/x^2$ and $z\equiv 1/(1+t^2)$ then (\ref{ss27})
can be rewritten%
$$
\Gamma _{lab}=\frac 16\int V_0(x)4t^3z^{5/2}\sum_{j=0}^\infty {{-\frac
12}\choose j}(-1)^jz^jdx=\frac 16\int V_0(x)4t^3z^2\sqrt{\frac z{1-z}}dx, 
$$
which implies 
\begin{equation}
\label{ss28}\Gamma _{lab}=\frac 16\int_{x_0^2}^1\frac{4x_0^2/x^2}{%
(1+x_0^2/x^2)^2}V_0(x)dx. 
\end{equation}
Simultaneously, since 
$$
\int_{x_0^2}^1V_0(x)dx=-\int_{x_0^2}^1xV_0^{\prime
}(x)dx=-\int_{x_0^2}^1xV_{-1}^{\prime }(x)\left( \frac x2+\frac{x_0^2}{2x}%
\right) dx 
$$
$$
=-\int_{x_0^2}^1V_{-1}^{\prime }(x)\left( \frac{x^2}2+\frac{x_0^2}2\right)
dx=\int_{x_0^2}^1V_{-1}(x)xdx, 
$$
the integral (\ref{ss24}) can be rewritten 
\begin{equation}
\label{ss29}\Gamma _{IMF}=\frac 16\int_{x_0^2}^1V_0(x)dx. 
\end{equation}
Let us express the last integral as%
$$
\int_{x_0^2}^1V_0(x)dx=\int_{x_0^2}^{x_0}V_0(x)dx+\int_{x_0}^1V_0(x)dx 
$$
and modify the first integral on r.h.s. using substitution $y=x_0^2/x$%
$$
\int_{x_0^2}^{x_0}V_0(x)dx=\int_{x_0}^1V_0\left( \frac{x_0^2}y\right) \frac{%
x_0^2}{y^2}dy. 
$$
Now let us recall the general shape of the functions (\ref{ss23}) obeying
Eq. (P3.24), which implies%
$$
V_0\left( \frac{x_0^2}y\right) =V_0(y), 
$$
therefore instead of (\ref{ss29}) one can write 
\begin{equation}
\label{ss30}\Gamma _{IMF}=\frac 16\int_{x_0}^1V_0(x)\left( \frac{x^2+x_0^2}{%
x^2}\right) dx. 
\end{equation}
Similar modification of Eq. (\ref{ss28}) gives 
\begin{equation}
\label{ss31}\Gamma _{lab}=\frac 16\int_{x_0}^1V_0(x)\left( \frac{4x_0^2}{%
x^2+x_0^2}\right) dx. 
\end{equation}
Obviously, both the integrals are equal for $V_0$ sharply peaked around $%
x=x_0$, but generally, for non static quarks 
\begin{equation}
\label{ss32}\Gamma _{lab}<\Gamma _{IMF}. 
\end{equation}

\begin{figure}
\begin{center}
\epsfig{file=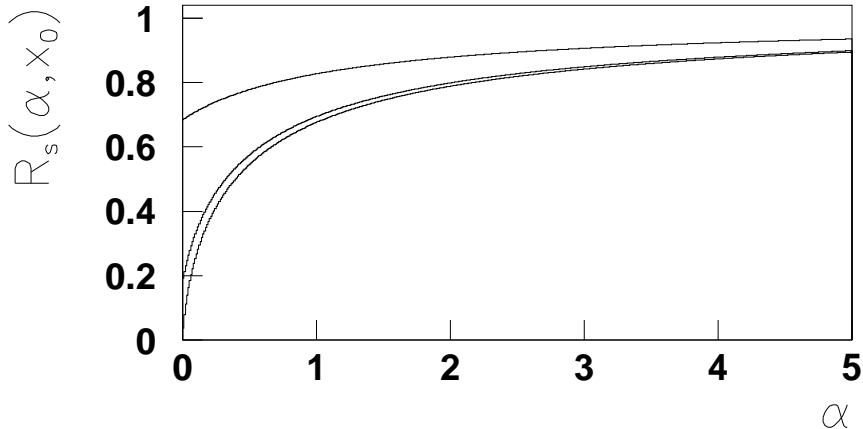, height=8cm}
\end{center}
\caption {Ratio $R_s $ plotted for values $x_0 $=0.2, 0.02, 0.0002 - 
in order from top to bottom.\label{prsa1}}
\end{figure}

What can our result (\ref{ss32}) mean quantitatively for the more realistic
scenario? In our discussion in \cite{zav} we have suggested the real
structure functions could be rather some superposition of our idealized ones
based on the single values of the effective mass $x_0=m/M.$ That means all
the relations involving the functions $V_j(x)\equiv V_j(x,x_0)$ should be
integrated over some distribution of the effective masses $\mu (x_0).$ But
at first, let us try to guess $V_0$, at least in the vicinity of $x_0$,
which is important for the integrals (\ref{ss30}) and (\ref{ss31}).
According to the Eq. (P3.20) for $x>x_0$ one can write 
\begin{equation}
\label{m1}xV_0^{\prime }(x)=-\frac M2P(p_0),\qquad p_0=\frac M2\left( x+
\frac{x_0^2}x\right) .
\end{equation}
Now, for $p_0$ close to $m$ let us parameterize the energy distribution by 
\begin{equation}
\label{m2}P(p_0)=\frac{\alpha \exp (\alpha )}m\exp \left( -\alpha \frac{p_0}%
m\right) ,
\end{equation}
which fulfills the normalization 
\begin{equation}
\label{m3}\int_m^\infty P(p_0)dp_0=1.
\end{equation}
Obviously, the distribution (\ref{m2}) means the average quark kinetic
energy equals to $m/\alpha $. Inserting (\ref{m2}) into (\ref{m1}) gives 
\begin{equation}
\label{m4}V_0^{\prime }(x)=-\frac{\alpha \exp (\alpha )}{2x_0x}\exp \left(
-\frac \alpha 2\left[ \frac x{x_0}+\frac{x_0}x\right] \right) .
\end{equation}
Let us note, for $\left| y\right| \ll 1$%
$$
(1+y)^a\approx \exp (ay), 
$$
therefore if we substitute the exponential function in (\ref{m4}) by 
\begin{equation}
\label{m4a}\exp \left( -\frac \alpha 2\left[ \frac x{x_0}+\frac{x_0}x\right]
\right) \sim \left[ \left( 1-x\right) \left( 1-\frac{x_0^2}x\right) \right]
^{\alpha /2x_0}\equiv f(x,x_0),\qquad x_0^2\leq x\leq 1,
\end{equation}
the resulting $V_0(x)$ will coincide with (\ref{m4}) in a vicinity of $x_0$,
but moreover will obey the global kinematical constraint outlined in Fig.
(P2). The ratio of integrals (\ref{ss31}) and (\ref{ss30}) calculated by
parts with the use of Eqs. (\ref{m4}) and (\ref{m4a}) gives 
\begin{equation}
\label{m5}R_s(\alpha ,x_0)\equiv \frac{\Gamma _{lab}}{\Gamma _{IMF}}=\frac{%
4\int_{x_0}^1x_0/x\left( \arctan [x/x_0]-\pi /4\right) f(x,x_0)dx}{%
\int_{x_0}^1\left( 1-x_0^2/x^2\right) f(x,x_0)dx},
\end{equation}
the results of the numerical computing are plotted in the Fig. \ref{prsa1}.
What do these curves mean? There are the two limiting cases:

a) The quarks are massive and static, i.e. $\alpha \rightarrow \infty ,$
then $R_s\rightarrow 1.$ It is the scenario in which both the approaches are
equivalent.

b) Both, the quark effective mass and $\alpha \ $$\rightarrow 0,$ but the
quark energy $\left\langle E_{kin}\right\rangle =m/\alpha >0,$ then $%
R_s\rightarrow 0.$ It is due to the fact that the massless fermions having
spin oriented always parallel to their momentum cannot contribute to spin
structure function of the system with the spherical phase space.

Obviously, the real case could be somewhere between both the extremes, i.e. $%
\alpha $ and $m$ should be the finite, positive quantities. The combination
of Eqs. (\ref{m5}) and (\ref{ss22}) gives 
\begin{equation}
\label{m6}\Gamma _{lab}=\frac 16R_s(\alpha ,x_0). 
\end{equation}
The comparison with the experimental value $\Gamma _{\exp }\simeq 0.13$
implies $R_s\simeq 0.78$, which according to the Fig. \ref{prsa1}
corresponds to $\alpha \simeq 2.$ Let us note, this result depends on $x_0$
rather slightly, therefore irrespective of the unknown distribution of
effective masses $\mu (x_0)$ we can conclude the following. If we accept the
quarks have on an average (over effective masses distribution) the mean
kinetic energy roughly equal to one half of the corresponding effective
mass, then within our approach, the experimental value $\Gamma _{\exp }$ is
compatible with the assumption that whole proton spin is carried by the
valence quarks.

\section{Summary and conclusion}

We have calculated the first moment $\Gamma _1$ of the proton spin structure
function in the approach which takes consistently into account the internal
motion of the quarks described by a spherical phase space. Simultaneously we
have done a comparison with the corresponding quantity deduced from the
standard IMF approach and came to the conclusion that the latter gives a
greater value $\Gamma _1.$ This difference is due to the fact, that the
standard approach is based on the approximation (P3.36), which effectively
suppresses the internal motion of quarks. On the other hand, in our
approach, the total quark energy is shared between the effective mass and
the kinetic energy, and correspondingly the resulting formula correctly
reflects the mass dependence of the structure function: $\Gamma _1$
continuously vanishes for massless quarks controlled by a spherical phase
space. Let us note, the quark intrinsic motion has been shown to reduce $%
\Gamma _1$ also in some another approaches \cite{bo1}-\cite{rit}.

Finally, we came to the conclusion that our $\Gamma _1$ calculated only from
the valence quarks contribution is compatible with the experimental data -
provided that their kinetic energies are on an average roughly equal to one
half of their effective mass. The application of the constrains due to axial
vector current operators on the spin contribution from different flavors can
somewhat change the parameter $\alpha $ to achieve an agreement with the
data. This question is studied and will be discussed in a next paper.

The corrections on $\Gamma _1$ suggested in this paper together with the
corrections on the distribution functions (P3.59) should be taken into
account for interpretation of the experimental data. At the same time it is
obvious the distribution of effective masses $\mu (x_0)$ is a quantity
requiring further study.



\end{document}